\documentclass[aps,prb,reprint,superscriptaddress]{revtex4-1}
\usepackage{amssymb}
\usepackage{amsfonts}
\usepackage{amsmath}
\usepackage{graphicx}%
\usepackage{color}
\usepackage[english]{babel}
\usepackage{lipsum}

\begin{document}


\title{Frequency Modulation and Voltage Locking of the Voltage Controlled Spin Oscillators (VCSOs)}


\author{Lang Zeng}
\email[]{zenglang@buaa.edu.cn}
\affiliation{Fert Beijing Institute, Beijing Advanced Innovation Center for Big Data and Brain Computing, Beihang University, Beijing, China, 100191}
\affiliation{School of Microelectronics, Beihang University, Beijing, China, 100191}
\affiliation{Hefei Innovation Research Institute, Beihang University, Hefei, Anhui, China, 230013}


\author{Hao-Hsuan Chen}
\thanks{Lang Zeng and Hao-Hsuan Chen contributed equally to this work.}
\affiliation{Fert Beijing Institute, Beijing Advanced Innovation Center for Big Data and Brain Computing, Beihang University, Beijing, China, 100191}
\affiliation{School of Microelectronics, Beihang University, Beijing, China, 100191}
\affiliation{Hefei Innovation Research Institute, Beihang University, Hefei, Anhui, China, 230013}

\author{Deming Zhang}
\affiliation{Fert Beijing Institute, Beijing Advanced Innovation Center for Big Data and Brain Computing, Beihang University, Beijing, China, 100191}
\affiliation{School of Electronic and Information Engineering, Beihang University, Beijing, China, 100191}

\author{Tianqi Gao}
\affiliation{Fert Beijing Institute, Beijing Advanced Innovation Center for Big Data and Brain Computing, Beihang University, Beijing, China, 100191}
\affiliation{School of Microelectronics, Beihang University, Beijing, China, 100191}

\author{Mingzhi Long}
\affiliation{Fert Beijing Institute, Beijing Advanced Innovation Center for Big Data and Brain Computing, Beihang University, Beijing, China, 100191}
\affiliation{School of Electronic and Information Engineering, Beihang University, Beijing, China, 100191}

\author{Youguang Zhang}
\affiliation{School of Electronic and Information Engineering, Beihang University, Beijing, China, 100191}

\author{Weisheng Zhao}
\email[]{weisheng.zhao@buaa.edu.cn}
\affiliation{Fert Beijing Institute, Beijing Advanced Innovation Center for Big Data and Brain Computing, Beihang University, Beijing, China, 100191}
\affiliation{School of Microelectronics, Beihang University, Beijing, China, 100191}


\date{\today}

\begin{abstract}
The oscillating frequency of typical Spin Torque Nano Oscillators (STNOs) can be modulated by injected DC current or bias magnetic field. And phase locking of STNOs to an external Radio Frequency (RF) signal can be imposed by AC current or RF bias magnetic field. However, in this study, we have proposed a Voltage Controlled Spin Oscillators (VCSOs) by introducing Voltage Controlled Magnetic Anisotropy (VCMA) effect. The oscillating frequency of VCSOs can be modulated by VCMA voltage as well as injected DC current. Furthermore, we have shown a novel locking mechanism caused by AC VCMA voltage. Both the frequency modulation and voltage locking mechanism are analyzed theoretically by Nonlinear Auto-oscillator theory and verified by numerical simulation. At last, we proposed that by utilizing negative capacitance material to enhance VCMA effect, the locking range for voltage locking can be expanded thus may lead to easy mutual synchronization of multiple VCSOs.
\end{abstract}

\pacs{}

\maketitle

\section{introduction}

Spin Torque Nano Oscillators (STNOs) are very promising devices which can be used to replace traditional Voltage Controlled Oscillator (VCO) in semiconductor chips~\cite{STNO1,STNO2,STNO3,STNO5}. Comparing with VCO, STNO has advantages of small footprint size ($\sim 100\ \rm{nm}$) and wide tunable frequency range ($\sim\rm{GHz}$). However, the STNO is suffering from low emitted power ($\sim \rm{\mu W}$) and high phase and amplitude noise\cite{STNO_Power1,STNO_Power2,STNO_noise1,STNO_noise2,STNO_noise3,STNO_noise4,STNO_noise5}. It is both theoretically predicted and experimentally demonstrated that for $N$ synchronized STNOs, the emitted power can be increased as $N^2$ while the frequency line width can be decreased as $N^{-2}$~\cite{STNO_syn1,STNO_syn2,STNO_syn3,STNO_syn4,STNO_syn5}. The synchronization of multiple STNOs can be performed through electrical connection~\cite{STNO_syn1,STNO_syn5,STNO_syn15}, spin wave propagation~\cite{STNO_syn7,STNO_syn10,STNO_syn11,STNO_syn12,STNO_syn13,STNO_syn14} as well as dipolar effect~\cite{STNO_syn6,STNO_syn8,STNO_syn9}. Among all these synchronization mechanisms, the most convenient one to be controlled and implemented in semiconductor chip is electrical connection. Since the low emitted power of single STNO, the mutual synchronization by electrical connection is very difficult to experimentally achieve~\cite{STNO_syn1,STNO_syn5}. The only example for such success is mutual synchronization of electrically connected two vortex STNOs because of large output power of vortex STNOs~\cite{STNO_syn15}. The oscillating frequency of vortex STNOs is only several hundreds of MHz which is far below frequency band used in modern communication system~\cite{Vortex}. The electrical synchronization is closely related to injection locking phenomenon in STNO~\cite{lockingrange1,lockingrange2,lockingrange3}. If the locking range can be expanded, it is postulated that the electrical synchronization can be made easily.

The stable rotation of magnetic moment in the free layer of STNOs is sustained by injected DC current which will induce Spin Transfer Torque (STT) effect. This implies that STNO is basically a current controlling device. Thus the way for electrical mutual synchronization of multiple STNOs is through injecting AC current~\cite{STNO_syn1,STNO_syn5,STNO_syn15}. However, the Voltage Controlled Oscillator composed by semiconductor devices using in chips are voltage controlling devices~\cite{VCO}. If the voltage can be introduced into the controlling of STNOs as an additional degree of freedom, not only the controlling of STNOs can be more versatile, but also novel locking and synchronization by applying AC voltage can be imaged.

The emitted power of STNOs is related to their magnetic configurations of free layer and reference layer. There are four kinds of different configurations~\cite{STNO3}: magnetic directions of both layers are in-plane (case 1), magnetic directions of both layers are perpendicular (out-of-plane) (case 2), magnetic direction of free layer is in-plane and that of reference layer is perpendicular (case 3) and vice versa (case 4). Among all of these four configurations, the case 1 and case 2 can not provide large magnetoresistance change thus lead to small emitted power while case 3 need another in-plane magnetization layer for signal readout. The configuration of case 4 can provide intrinsically large magnetoresistance change which is beneficial for power emission~\cite{Kubota1,Kubota2,Kubota3,Kubota4,Kubota5,Kubota6}. Experimentally the high emitted power of STNO with case 4 configuration is demonstrated. The $\rm{\mu W}$ emission power is almost the highest one reported in the literature for single STNO device excluding vortex type~\cite{Kubota3}. The oscillating frequency of the STNO with in-plane reference layer and perpendicular free layer can be controlled by injected DC current, bias magnetic field as well as Perpendicular Magnetic Anisotropy (PMA) field of the free layer. The PMA field of the free layer can be modulated by Voltage Controlled Magnetic Anisotropy (VCMA) effect with an applying voltage~\cite{VCMA1,VCMA2,VCMA3,VCMA4,VCMA5}. This inspires us how to make the STNO device to be voltage controlled.

In this work, we proposed Voltage Controlled Spin Oscillators (VCSOs) which has the voltage as an additional degree of freedom by introducing VCMA effect for the controlling of oscillating. The frequency modulation and the locking phenomenon by both injected current and applied voltage are theoretically analyzed with Nonlinear Auto-Oscillator theory~\cite{nonlinear1,nonlinear2}. In contrast with current locking mechanism, the voltage locking mechanism can be explained by the tilted oscillation axis. The locking range of the voltage locking is comparable to that of the current locking. However, we furthermore proposed that a negative capacitance layer can be added in the stack which can enhance the VCMA effect by tens of times~\cite{zeng1,zeng2}. This will make the locking range of the voltage locking expand correspondingly and it implies an easy and novel way for electrical mutual synchronization of multiple VCSOs.

\section{Frequency Modulation of Proposed VCSO}
\subsection{Device Structure of Proposed VCSO}

\begin{figure}
\includegraphics[width=0.5\textwidth]{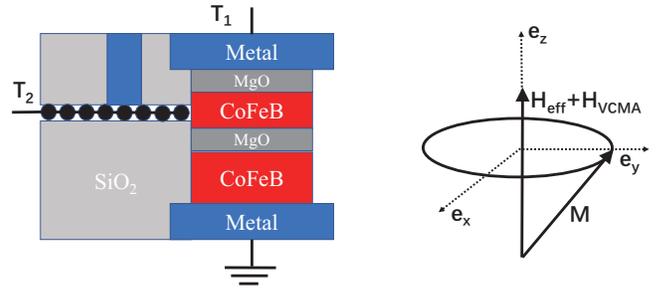}
\caption{\label{VCSO} The structure of the proposed Voltage Controlled Spin Oscillator (VCSO). From bottom to up it is bottom contact/thick reference layer/thin MgO/thin free layer/thick MgO/up contact. The magnetization direction of the thick reference layer is in-plane while the magnetization direction of the thin free layer is out-of-plane (perpendicular). When applying a voltage between Terminal T$_1$ and T$_2$, the VCMA effect will modify the PMA field of the free layer, thus tune the oscillation frequency of the free layer.}
\end{figure}

The device we proposed is illustrated in Fig.~\ref{VCSO}. The proposed device has three terminals: T$_1$, T$_2$ and $T_3$ (Ground). The terminal T$_2$ connects to a thin CoFeB layer which holds perpendicular magnetic direction because of interfacial PMA effect and serves as free layer. The terminal T$_3$ connects to a thick CoFeB layer which holds in-plane magnetic direction and serves as reference layer. The MgO insulating layer between free layer and reference layer is about 0.8 nm thick to keep the RA product of Magnetic Tunneling Junction (MTJ) small. The MgO layer between terminal T$_1$ and free CoFeB layer is about 1.2 nm thick to make the resistance between terminal T$_2$ and T$_1$ large. Applying a voltage between terminal T$_2$ and T$_1$, the current can be negligible since large resistance. However, the VCMA effect will take place and the applying voltage will modify the PMA field of the free layer.

As shown in Fig.~\ref{VCSO}, the perpendicular direction is defined as z axis in our study while the in-plane direction is x and y axis. When there is injected DC current between terminal T$_2$ and T$_3$, the spin torque exerted on the magnetic moment of the free layer will make it rotate around the z axis. From previous analytical solution, it can seen that the oscillating frequency of STNO is related by the external magnetic field, the PMA field and the injected DC current. Since the applying VCMA voltage can modify the PMA field of the free layer, the oscillating frequency can be tuned by the applying VCMA voltage. This makes our proposed device is voltage controllable and we named it Voltage Controlled Spin Oscillator (VCSO).

\subsection{Theoretical Analysis on Frequency Modulation}

The analytical solution for the frequency modulation of STNO with perpendicular free layer and in-plane reference layer is already derived by energy balance theory and verified with experimental results. However in energy balance theory the energy change due to spin torque term and damping term is averaged over one rotation period, thus some information about the oscillating dynamics is lost. Here in this study, the phase locking phenomenon will be explored. We expanded Nonlinear Auto-oscillator theory developed by J. V. Kim, V. Tiberkevich and A. Slavin for our study~\cite{nonlinear1,nonlinear2}. The Nonlinear Auto-oscillator theory will provide unified and powerful framework for the analysis of VCSO.

The LLG equation which describes the magnetic dynamics in the free CoFeB layer is as follows
\begin{equation}
\frac{\partial m}{\partial t} = -\gamma(m\times H_{\text{eff}})+\alpha(m\times\frac{\partial m}{\partial t})-\sigma J m\times(m\times e_{\text{p}})
\label{LLG1}
\end{equation}
where $m$ is the magnetic direction in the free layer, $H_{\text{eff}}$ is the effective field of the free layer, $\alpha$ is the Gilbert damping constant, $\gamma$ is the gyromagnetic ratio, $\sigma$ is the spin torque efficient, $J$ is the injected current density and $e_{\text{p}}$ is the vector describing the magnetic direction in the reference layer.

Eq.~\ref{LLG1} can be rearranged as
\begin{equation}
\begin{aligned}
\frac{\partial m}{\partial t} = &-\frac{\gamma}{1+\alpha^2}(m\times H_{\text{eff}})-\alpha\frac{\gamma}{1+\alpha^2}(m\times(m\times H_{\text{eff}}))\\
&-\frac{\sigma J}{1+\alpha^2}[m\times(m\times e_{\text{p}})-\alpha(m\times e_{\text{p}})]
\label{LLG2}
\end{aligned}
\end{equation}
Since Gilbert damping constant is very small, Eq.~\ref{LLG2} can be approximated by
\begin{equation}
\begin{aligned}
\frac{\partial m}{\partial t} = &-\gamma(m\times H_{\text{eff}})-\alpha\gamma(m\times(m\times H_{\text{eff}}))\\
&-\sigma J (m\times(m\times e_{\text{p}}))
\end{aligned}
\label{LLG3}
\end{equation}

In our VCSOs, the magnetic direction of the reference layer is in-plane and it can be assumed in the x axis.
\begin{equation}
e_{\text{p}} =  \left(\begin{array}{c}
P_x\\
0\\
0\\
\end{array}
\right)
\end{equation}
If the applied VCMA voltage is zero, then the effective field $H_{\text{eff}}$ can be written as
\begin{equation}
H_{\text{eff}} =  \left(\begin{array}{c}
0\\
0\\
H_{\text{appl}}+(H_k-4\pi M_s)m_z\\
\end{array}
\right)
\end{equation}
Both of the external magnetic field $H_\text{appl}$ and the effective PMA field $(H_k-4\pi M_s)m_z$ are along z axis.

The LLG equation Eq.~\ref{LLG3} is expanded to be
\begin{widetext} 
\begin{eqnarray} 
\begin{aligned}
&\left(\begin{array}{c}
\dot{m_x}\\
\dot{m_y}\\
\dot{m_z}\\
\end{array}
\right)
=  -\gamma\left(\begin{array}{c}
m_x\\
m_y\\
m_z\\
\end{array}
\right)\times\left(\begin{array}{c}
0\\
0\\
H_{\text{appl}}+(H_k-4\pi M_s)m_z\\
\end{array}
\right)\\
&-\alpha\gamma\left(\begin{array}{c}
m_x\\
m_y\\
m_z\\
\end{array}
\right)\times\left[\left(\begin{array}{c}
m_x\\
m_y\\
m_z\\
\end{array}
\right)\times\left(\begin{array}{c}
0\\
0\\
H_{\text{appl}}+(H_k-4\pi M_s)m_z\\
\end{array}
\right)\right]\\
&-\sigma J \left(\begin{array}{c}
m_x\\
m_y\\
m_z\\
\end{array}
\right)\times\left[\left(\begin{array}{c}
m_x\\
m_y\\
m_z\\
\end{array}
\right)\times\left(\begin{array}{c}
P_x\\
0\\
0\\
\end{array}
\right)\right]\\
\end{aligned}
\end{eqnarray}
\end{widetext}

A new variable $a$ defined by
\begin{equation}
a = \frac{m_x-i m_y}{\sqrt{2(1+m_z)}}
\end{equation}
is introduced. With this definition, $m_x$, $m_y$ and $m_z$ can be written as
\begin{equation}
\begin{aligned}
&m_x = (a+a^*)\sqrt{1-|a|^2}\\
&m_y = i(a-a^*)\sqrt{1-|a|^2}\\
&m_z = 1-2|a|^2\\
\end{aligned}
\end{equation}
With this transformation, the precessional term $-\gamma(m\times H_\text{eff})$ is derived as
\begin{equation}
\begin{aligned}
\left.\frac{\partial a}{\partial t}\right|_\text{Precession} &= -i\gamma \left[\begin{array}{c}
(H_{\text{appl}}+H_k-4\pi M_s)a\\
-2(H_k-4\pi M_s)a^2a^*\\
\end{array}
\right]\\
&= -i\gamma \left[\begin{array}{c}
(H_{\text{appl}}+H_k-4\pi M_s)a\\
-2(H_k-4\pi M_s)pa\\
\end{array}
\right]\\
\end{aligned}
\end{equation}
where $p=|a|^2$ is the auto-oscillation power.

Similarly, the damping term $-\alpha\gamma(m\times(m\times H_{\text{eff}}))$ is written as
\begin{equation}
\begin{aligned}
\left.\frac{\partial a}{\partial t}\right|_\text{Damp} &= \alpha\gamma\left[\begin{array}{c}
-(H_{\text{appl}}+H_k-4\pi M_s)a\\
(H_{\text{appl}}+3(H_k-4\pi M_s))a^2a^*\\
-2(H_k-4\pi M_s)a^3a^{*2}\\
\end{array}
\right]\\
&= \alpha\gamma\left[\begin{array}{c}
-(H_{\text{appl}}+H_k-4\pi M_s)a\\
(H_{\text{appl}}+3(H_k-4\pi M_s))pa\\
-2(H_k-4\pi M_s)p^2a\\
\end{array}
\right]\\
\end{aligned}
\end{equation}
As we can see, both of the precessional term and damping term can be written as polynomial expansion of $a$ and $p$. However, the STT term $-\sigma J (m\times(m\times e_{\text{p}}))$ is transformed to
\begin{equation}
\left.\frac{\partial a}{\partial t}\right|_\text{STT} = \frac{\sigma J }{4\sqrt{1-aa^*}}\left[\begin{array}{c}
2\\
-3aa^*\\
-3a^2\\
2a^2a^{*2}\\
2a^3a^*\\
\end{array}
\right]
\label{STT_all}
\end{equation}
Since there is $\sqrt{1-aa^*}$ in the denominator, direct following analysis in the framework of Nonlinear auto-oscillator is not possible.

The spin torque strength $\sigma$ is defined by
\begin{equation}
\begin{aligned}
\sigma& = \frac{\gamma\hbar\eta}{2e(1+\lambda m_x)M_s Sd}\\
&=\frac{\gamma\sigma^{\prime}}{(1+\lambda m_x)}\\
\end{aligned}
\end{equation}
where $\sigma^{\prime}$ is
\begin{equation}
\sigma^{\prime} = \frac{\hbar\eta}{2eM_s Sd}
\end{equation}
$\lambda$ is a small value, and for not very large oscillation angle $m_x\ll 1$, thus the spin torque coefficient can be in Taylor expansion form as
\begin{equation}
\sigma = \gamma\sigma^{\prime}(1-\lambda m_x+\lambda^2 m_x^2-\lambda^3 m_x^3+\text{O}(\lambda^4 m_x^4))
\end{equation}
Also for not very large angle oscillation, $p\ll 1$. Thus the denominator $\sqrt{1-aa^*}$ can be Taylor expanded also
\begin{equation}
\sqrt{1-aa^*} = 1-\frac{aa^*}{2}-\frac{(aa^*)^2}{8}-\frac{(aa^*)^3}{16}+\text{O}((aa^*)^4)
\end{equation}

With these two Taylor expansions, the STT term can be written to the order of $p^2$
\begin{equation}
\left.\frac{\partial a}{\partial t}\right|_\text{STT} = \frac{\gamma\sigma^{\prime}\lambda J}{2}(1-(3-3\lambda^2)p+(2-12\lambda^2)p^2)a 
\label{p2} 
\end{equation}
and to the order of $p^3$
\begin{equation}
\left.\frac{\partial a}{\partial t}\right|_\text{STT} = \frac{\gamma\sigma^{\prime}\lambda J}{2}(1-(3-3\lambda^2)p+(2-12\lambda^2)p^2+15\lambda^2p^3)a  
\label{p3}
\end{equation}
Now all of the terms in LLG Equation Eq.~\ref{LLG3} is the function of $p$ and $a$. Keep in mind that in Eq.~\ref{p2} and \ref{p3}, only conservative terms around the oscillation orbit are written. For the derivation of oscillation frequency, only conservative terms are reqiured. However, for the exploration of phase locking phenomenon, non-conservative terms in Eq.~\ref{STT_all}. So we can apply similar analysis for the frequency modulation and locking range in the framework of Nonlinear Auto-oscillator theory. It is worth noting that the calculation for Eq.~\ref{p2} and \ref{p3} is tedious and complicated, we invented a new method call Virtual Perpendicular System for the  fast theoretical derivation. The details are presented in  Appendix.~\ref{VPS}.

After the above calculation, the oscillation dynamics of the proposed VCSO can be described by the general nonlinear oscillator model
\begin{equation}
\frac{da}{dt}+i\omega(p)a+\Gamma_+(p)a-\Gamma_-(p)a=0
\label{KTS}
\end{equation}
where $a$ is the the complex amplitude of the auto-oscillation, and $p$ is the power of the auto-oscillation and $\phi = \text{arg}(a)$ is the phase of the auto-oscillation. $\omega(p)$ is the oscillating frequency of the auto-oscillator, and the detail expression for it is written as
\begin{equation}
\omega(p) = \omega_0+Np
\label{omega}
\end{equation}
where $N$ is the coefficient of the nonlinear frequency shift for the normally magnetized film in the free CoFeB layer. $N$ can be written as
\begin{equation}
N = -2\omega_k
\end{equation}
In the above two equations, $\omega_0$ and $\omega_k$ is
\begin{equation}
\begin{aligned}
&\omega_0 = \gamma(H_{\text{appl}}+H_k-4\pi M_s)\\
&\omega_k = \gamma(H_k-4\pi M_s)\\
\label{omega_0}
\end{aligned}
\end{equation}

The $\Gamma_+(p)$ in Eq.~\ref{KTS} is the damping rate for the natural energy dissipation of the auto-oscillation and it comes from the damping torque. The exact form of $\Gamma_+(p)$ is
\begin{equation}
\Gamma_+(p) = \Gamma_G(1+Qp+Q^{\prime}p^2)
\end{equation}
where
\begin{equation}
\begin{aligned}
&\Gamma_G = \alpha\omega_0\\
&Q = -(1+2\omega_k/\omega_0)\\
&Q^{\prime} = 2\omega_k/\omega_0\\
\label{Gamma_+}
\end{aligned}
\end{equation}
For physically-relevant range of complex amplitudes $a$, the damping term $\Gamma_+(p)$ is positive thus $\Gamma_+(p)$ describes the reduction of the spin wave amplitude $a$ with time.

Contrary to damping term $\Gamma_+(p)$, the STT term $\Gamma_-(p)$ comes from the STT torque of the injected current and it describes the effective action of the external energy source. To calculate $\Gamma_-(p)$, as shown in the above Eq.~\ref{p2} and \ref{p3}, extensions to the Nonlinear Auto-oscillator theory should be made. After Taylor expansion, the expression for $\Gamma_-(p)$ to the order of $p^3$ is
\begin{equation}
\begin{aligned}
\Gamma_-(p) = &\frac{\gamma\sigma^{\prime}\lambda J}{2}(1-(3-3\lambda^2)p\\
&+(2-12\lambda^2+10\lambda^4)p^2+(15\lambda^2-50\lambda^4)p^3)\\
\end{aligned}
\end{equation}

The condition for the threshold injected current of the appearance of an auto-oscillation having a nonzero power is
\begin{equation}
\Gamma_+(0) = \Gamma_-(0)
\label{threshold}
\end{equation}
From Eq.~\ref{threshold}, the threshold current for the onset of auto-oscillation is
\begin{equation}
\begin{aligned}
I_{\text{th}} &= \frac{2\Gamma_G}{\gamma\sigma^{\prime}\lambda}\\
&= \frac{4\alpha e M_sSd}{\hbar\eta\lambda}(H_{\text{appl}}+H_k-4\pi M_s)\\
\end{aligned}
\end{equation}
This threshold current $I_\text{th}$ agrees exactly with the results obtained from the energy balance theory.

For the stability of the stationary solution of Eq.~\ref{KTS}, three more new variables can be defined as
\begin{equation}
\begin{aligned}
&G_+ = \frac{d\Gamma_+(p)}{dp}\\
&G_- = \frac{d\Gamma_-(p)}{dp}\\
&\Gamma_{p} = (G_+-G_-)p_0\\
\end{aligned}
\end{equation}
where the derivatives are calculated at the stationary point $p=p_0$. $\Gamma_p$ is the damping rate for small power deviations $\delta p$ at $p_0$. For the stationary solution $p=p_0$ is stable, it is required that
\begin{equation}
\Gamma_p > 0
\label{stationary1}
\end{equation}
If both of the damping term $\Gamma_+(p)$ and the STT term $\Gamma_-(p)$ is expanded to the order of $p$, the condition for stationary stable oscillation Eq.~\ref{stationary1} can be written
\begin{equation}
Q > -(3-3\lambda^2)
\end{equation}
Substituting $Q$ with Eq.~\ref{omega_0} and \ref{Gamma_+}, the condition for stationary stable oscillation is got
\begin{equation}
H_{\text{appl}} > 0
\end{equation}
The requirement for an external magnetic field in the z axis to sustain a stable oscillation which is the main conclusion of Ref.~33 is derived by us using Nonlinear Auto-oscillator theory.

For the stable stationary oscillation condition Eq.~\ref{stationary1}, if the damping term $\Gamma_+(p)$ and the STT term $\Gamma_-(p)$ is expanded to the order of $p^2$, then we can get
\begin{equation}
\begin{aligned}
1+Qp+Q^{\prime}p^2&=1-(3-3\lambda^2)p+(2-12\lambda^2)p^2\\
Q+Q^{\prime}p&=-(3-3\lambda^2)+(2-12\lambda^2)p\\
p&=\frac{Q+3-3\lambda^2}{2-12\lambda^2-Q^{\prime}}\\
\end{aligned}
\end{equation}
For stable stationary oscillation, $p>0$ should be fulfilled and after calculation it is got that
\begin{equation}
H_{\text{appl}} > \frac{3\lambda^2}{2-3\lambda^2}(H_k-4\pi M_s)
\end{equation}
This result means that there exists a minimum external magnetic field in the z axis to make the stable stationary oscillation possible. It is also worth noting that this is the main conclusion of Ref.~32 derived by energy balance theory.

For the calculation of oscillation frequency for any specific inject DC current, first the stationary oscillation power $p$ will be calculated from
\begin{equation}
\Gamma_+(p) = \Gamma_-(p)
\label{oscillation_frequency}
\end{equation}
Then Eq.~\ref{omega} will be used to calculate the oscillation frequency. It should be noted that for $\Gamma_-(p)$ expanded with the order of $p^2$ and $p^3$, there will be two or three roots for the solution of Eq.~\ref{oscillation_frequency}. However, since
\begin{equation}
\begin{aligned}
p&=&aa^*\\
&=& \frac{1-m_z}{2}
\end{aligned}
\end{equation}
So $0<p<1/2$. There will be only one root in this range. This implies our oscillation is single mode.

\begin{figure}
\includegraphics[width=0.45\textwidth]{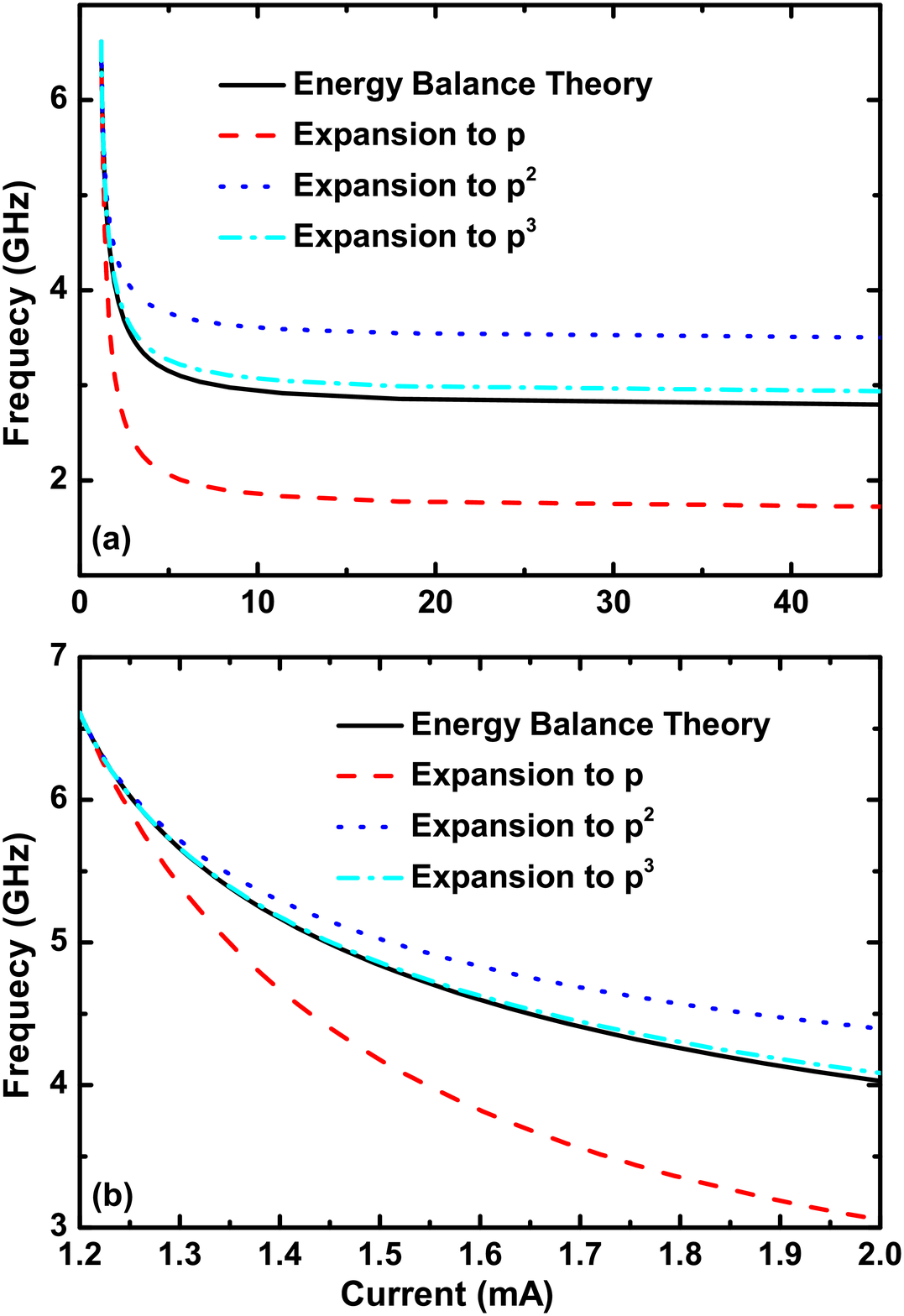}
\caption{\label{fig1} (a) and (b) Current modulated frequency calculated by Energy Balance Theory and our extended Nonlinear Auto Oscillator Theory which is expanded to orders of $p$, $p^2$ and $p^3$, respectively.}
\end{figure}

The analytical solution for the oscillation frequency with varied injected DC current is shown in Fig.~\ref{fig1}. We expanded the STT term $\Gamma_-(p)$ to the order of $p$, $p^2$ and $p^3$, and compared the results with that got by energy balance theory. Fig.~\ref{fig1}(a) shows the oscillation frequency for injected current in the range of 0-45 mA, and Fig.~\ref{fig1}(b) shows the oscillation frequency for injected current in the range of 1.2-2.0 mA. It is can be observed that there are large deviations when only expanded to the order of $p$ and $p^2$. However, when expanded to the order of $p^3$, the oscillation frequency calculated from Eq.~\ref{oscillation_frequency} is very close to the oscillation frequency derived from energy balance theory, especially when inject current is small.

Besides oscillation power $p_0$, there are two important variables defined as
\begin{equation}
\Gamma_p = (G_+-G_-)p_0
\end{equation}
and
\begin{equation}
\nu = \frac{N}{G_+-G_-}
\end{equation}
$\nu$ is related to the influence of the nonlinear frequency shift $N$ in comparison with the nonlinear damping coefficients $G_+$ and $G_-$. For conventional oscillators $\nu\ll 1$, the oscillation power $p$ has negligible effect on the oscillation frequency $\omega$. In contrary, for typical spin torque oscillators $\nu\gg 1$, thus spin torque oscillators is strong nonlinear oscillator.

\begin{figure}
\includegraphics[width=0.45\textwidth]{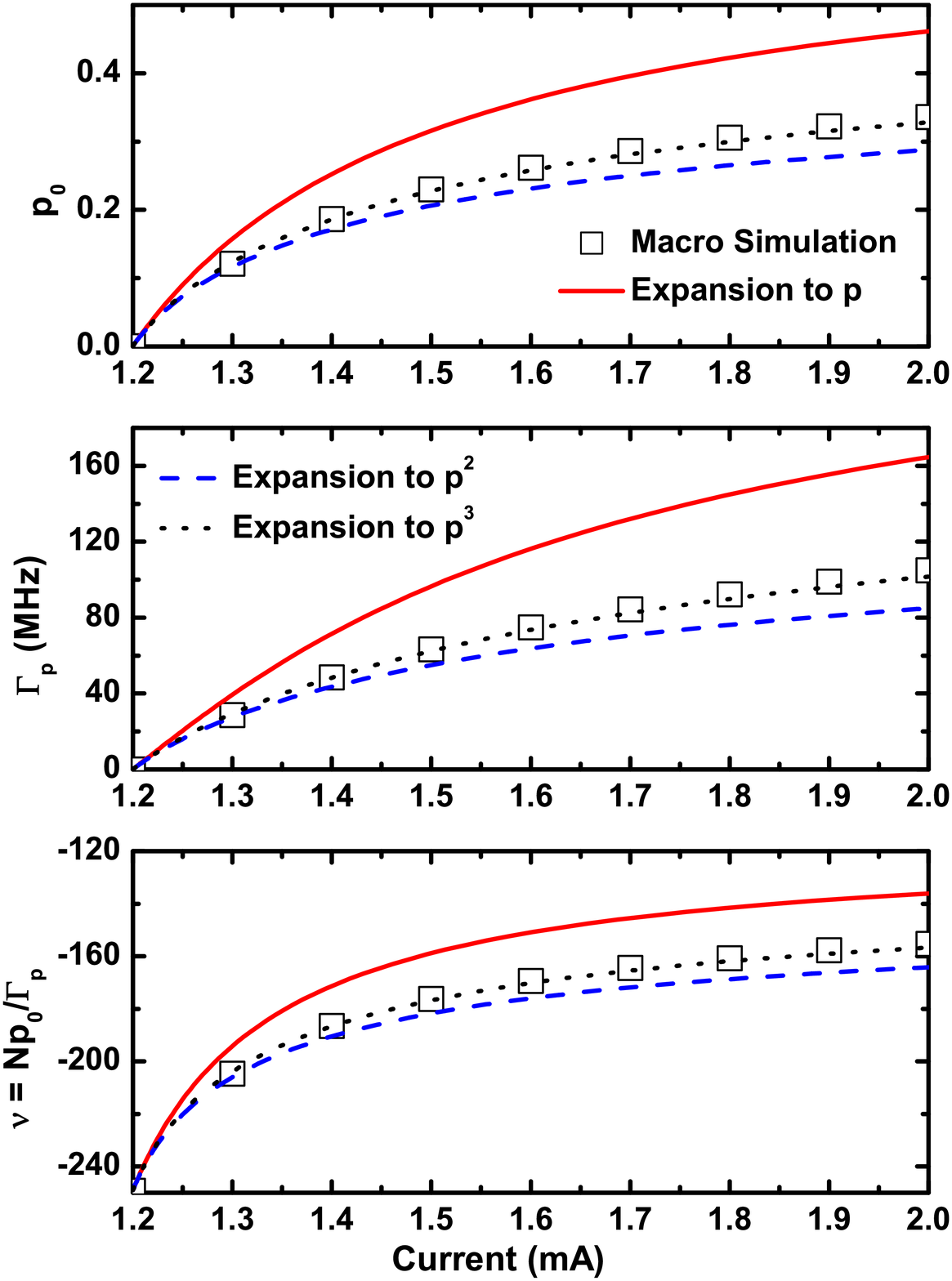}
\caption{\label{fig2} (a), (b) and (c) Spin wave power $p_0$, amplitude relaxation rate $\Gamma_p$ and non-linearity coefficient $\nu$ with varied injected DC current simulated by macro simulation and calculated by our extended Nonlinear Auto Oscillator Theory which is expanded to orders of $p$, $p^2$ and $p^3$, respectively.}
\end{figure}

The $p_0$, $\Gamma_p$ and $\nu$ calculated from Eq.~\ref{oscillation_frequency} and extracted from macro simulation are plotted in Fig.~\ref{fig2}. It is also observed that expansion to the order of $p^3$ can give us very accurate solutions agreeing well with macro simulation results.

\begin{figure}
\includegraphics[width=0.45\textwidth]{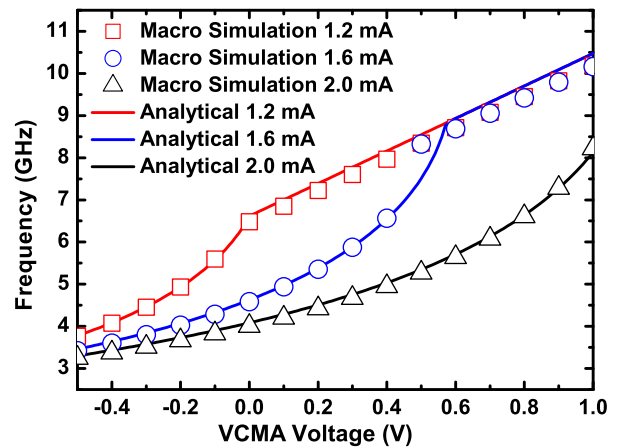}
\caption{\label{FvsV}The oscillation frequency can be modulated by applying VCMA voltage. The analytical results of the oscillation frequency for expansion to $p^3$ agree very well with that of macro simulation results.}
\end{figure}

When there applies VCMA voltage, the effective field can be written as
\begin{equation}
H_{\text{eff}} =  \left(\begin{array}{c}
0\\
0\\
H_{\text{appl}}+(H_k-4\pi M_s)(1+V_\text{VCMA}/V_\text{ref})m_z\\
\end{array}
\right)
\end{equation}
where $V_\text{VCMA}$ is the applied VCMA voltage, and $V_\text{ref}$ which means an applied VCMA voltage of $-V_\text{ref}$ will fully compensate the intrinsic PMA field of the free CoFeB layer. The oscillation frequency with varied applied VCMA voltage for injected DC current of 1.2, 1.6 and 2.0 mA respectively is shown in Fig.~\ref{FvsV}. The analytical results are calculated with the order of $p^3$. The analytical results agree very well with that of macro simulation results.

\begin{figure}
\includegraphics[width=0.45\textwidth]{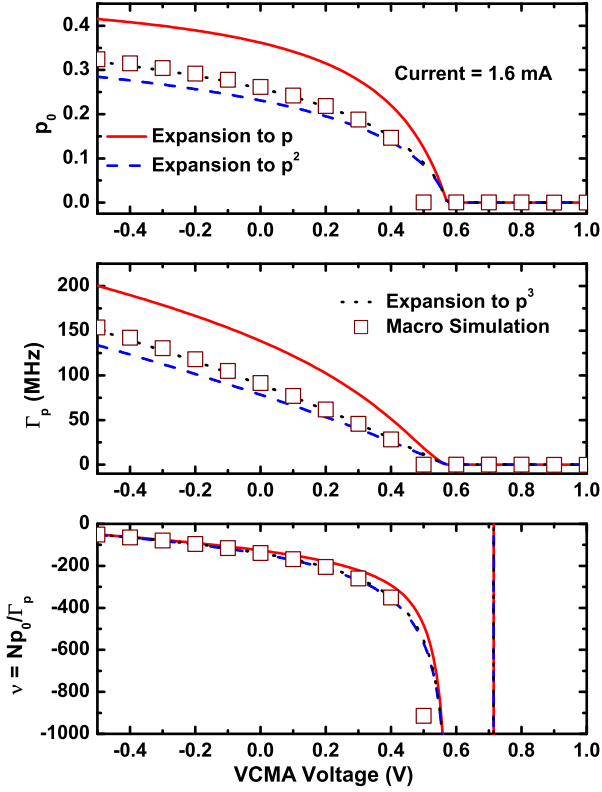}
\caption{\label{fig4} (a), (b) and (c) Spin wave power $p_0$, amplitude relaxation rate $\Gamma_p$ and non-linearity coefficient $\nu$ with varied VCMA voltage for injected DC current is 1.6 mA simulated by macro simulation and calculated by our extended Nonlinear Auto Oscillator Theory which is expanded to orders of $p$, $p^2$ and $p^3$, respectively.}
\end{figure}

The variables of $p_0$, $\Gamma_p$ and $\nu$ calculated analytically and extracted from macro simulation are plotted in Fig.~\ref{fig4}. The expansion to the order of $p^3$ can give us very accurate results for these three variables.

In this section, we derived frequency modulation by injected DC current and applied VCMA voltage under the framework of the Nonlinear Auto-oscillator theory. We got the same results as the analytical solution derived by energy balance theory, and repeated several main conclusions in the literature. Also our theoretical results agree very well with the macro simulation results. For the macro simulation, all of the parameters used are list in Table.~\ref{table1} extracted from experimental work~\cite{Kubota1,Kubota2,Kubota3,Kubota4,Kubota5,Kubota6}.

\begin{table}
\caption{\label{table1}Parameters used in the macro simulation.}
\begin{tabular}{cc}
\hline \hline
Parameter&Value\\
\hline
$M_s$&1313 emu/cm$^3$\\
$H_k$&17.9 kOe\\
$H_\text{appl}$&1.0 kOe\\
$S$&$\pi \times 50 \times 50$ nm$^2$\\
$d$&2 nm\\
$\gamma$&17.32 MHz/Oe\\
$\alpha$&0.005\\
$\eta$&0.33\\
$\lambda$&0.38\\
$V_\text{ref}$&1 V\\
\hline \hline
\end{tabular}
\end{table}

\section{Current and Voltage Locking of Proposed VCSO}

In the previous section, the Nonlinear Auto-oscillator theory is employed to analyze frequency modulation by injected DC current and applied VCMA voltage. Same results also can be got by energy balance theory. In this section, we will use Nonlinear Auto-oscillator theory to analyze phase locking phenomenon caused by RF current as well as RF voltage. Such theoretical derivations cannot be performed by energy balance theory.

\subsection{Theretical Analysis on Current Locking of VCSO\label{current_section}}

For phase locking caused by injected RF current, the injected current can be described
\begin{equation}
J=J_\text{DC}(1+\epsilon\cos(\phi_e))
\label{AC_current}
\end{equation}
where $\phi_e$ is the phase of the external injected RF current, $J_\text{DC}$ is the amplitude of the injected DC current and $\epsilon$ is the amplitude ratio of the external injected RF current and DC current. Eq.~\ref{AC_current} should not be substituted into Eq.~\ref{p2} or \ref{p3} since only conservative terms are kept thus no phase locking will be observed. Instead Eq.~\ref{AC_current} should be substituted into Eq.~\ref{STT_all}, and keep all of the coupling factors that allow for resonant with external RF current. After some tedious algebra calculation, we got
\begin{equation}
\begin{aligned}
\left.\frac{\partial a}{\partial t}\right|_\text{STT} = &\frac{\gamma\sigma^{\prime}\lambda J_\text{DC}}{2}[(1-(3-3\lambda^2)p+(2-12\lambda^2)p^2)a\\
&+\frac{1}{2\lambda}\epsilon\cos(\phi_e)(1+2\lambda^2a\cos(\phi)\sqrt{1-p})(2-3p)]\\
\end{aligned}
\label{current_locking1}
\end{equation}
where $\phi$ is the phase of the spin torque oscillator. Actually, the complex amplitude of the auto-oscillation $a$ can be written as
\begin{equation}
a = \sqrt{p}e^{i\phi}
\end{equation}
Thus we can write equations as follows
\begin{equation}
a^*\left(\frac{da}{dt}\right)=\frac{1}{2}\frac{dp}{dt}+ip\frac{d\phi}{dt}
\end{equation}
Multiple $a^*$ in the left of Eq.~\ref{current_locking1}, we can get
\begin{equation}
\begin{aligned}
a^*\left(\frac{da}{dt}\right)_{\text{STT}}=&\Gamma_-(p)p+\frac{\gamma\sigma^{\prime} J_\text{DC}}{4}\epsilon\cos(\phi_e)*\\
&(\sqrt{p}(\cos(\phi)-i\sin(\phi))\\
&+2\lambda^2p\cos(\phi)\sqrt{1-p})(2-3p)\\
\label{current_locking2}
\end{aligned}
\end{equation}

For the study of phase locking phenomenon with external RF current, the phase difference $\psi$ is introduced
\begin{equation}
\psi = \phi_e +\phi
\end{equation}
where the phase of the spin torque oscillator can be written as
\begin{equation}
\phi = -\omega t+\phi_0
\end{equation}
Noting since there is plus sign in the definition of phase difference, thus there should be minus sign in the definition of phase of the spin torque oscillator. Substituting $\psi$ into Eq.~\ref{current_locking2}, we can get
\begin{equation}
\begin{aligned}
a^*\left(\frac{da}{dt}\right)_{\text{STT}}=&\Gamma_-(p)p+\frac{\gamma\sigma^{\prime} J_\text{DC}}{4}\epsilon*\\
&(\frac{1}{2}\sqrt{p}(\cos(\psi)-i\sin(\psi))\\
&+\lambda^2p\cos(\psi)\sqrt{1-p})(2-3p)\\
\end{aligned}
\label{current_locking3}
\end{equation}
Rearranged Eq.~\ref{current_locking3} with oscillation power $p$ and oscillation phase $\phi$
\begin{equation}
\begin{aligned}
\frac{dp}{dt}=&-2(\Gamma_+(p)-\Gamma_-(p))p+\frac{\gamma\sigma^{\prime} J}{2}\epsilon*\\
&(\frac{1}{2}\sqrt{p}\cos(\psi)+\lambda^2p\cos(\psi)\sqrt{1-p})(2-3p)\\
\frac{d\phi}{dt}=&-(\omega_0+Np)-\gamma\sigma^{\prime} J\epsilon \frac{1}{8\sqrt{p}}\sin(\psi)(2-3p)\\
\end{aligned}
\label{current_locking4}
\end{equation}
Eq.~\ref{current_locking4} is derived by STT term expanded to the order of $p^2$. If the STT term just is expanded to the order of $p$, a simpler equation can be got
\begin{equation}
\begin{aligned}
\frac{dp}{dt}=&-2(\Gamma_+(p)-\Gamma_-(p))p+\frac{\gamma\sigma^{\prime} J}{2}\epsilon\sqrt{p}\cos(\psi)\\
\frac{d\phi}{dt}=&-(\omega_0+Np)-\gamma\sigma^{\prime} J\epsilon \frac{1}{4\sqrt{p}}\sin(\psi)\\
\end{aligned}
\label{current_locking5}
\end{equation}

With an injected RF current, the stationary oscillation power will have small deviation from its free-running oscillation power $p_0$
\begin{equation}
p_s = p_0 +\delta p_s
\end{equation}
where $p_s$ is the stationary forced power which is the stationary power under the impact of the injected RF current, and $\delta p_s$ is the forced power deviations due to injected RF current. Here we will use Eq.~\ref{current_locking5} corresponding to the STT term expanded to the order of $p$  to derive phase locking range and phase difference, and give the phase locking range and phase difference when the STT term is expanded to the order of $p^2$ directly. Substituting $p_s$ back into Eq.~\ref{current_locking5}, and use stationary stable oscillation conditions of $p_0$
\begin{equation}
0=-2(\Gamma_+(p)-\Gamma_-(p))(p_0+\delta p_s)+\frac{\gamma\sigma^{\prime} J}{2}\epsilon\sqrt{(p_0+\delta p_s)}\cos(\psi)\\
\label{current_locking6}
\end{equation}
Eq.~\ref{current_locking6} can be further simplified to be
\begin{equation}
0=-2\Gamma_p(p_0)\delta p_s+\frac{\gamma\sigma^{\prime} J}{2}\epsilon\sqrt{p_0}\cos(\psi)\\
\end{equation}
Thus the small power deviation $\delta p_s$ can be computed
\begin{equation}
\frac{\delta p_s}{p_0} = \frac{\gamma\sigma^{\prime} J}{4\Gamma_p(p_0)\sqrt{p_0}}\epsilon\cos(\psi)\\
\end{equation}
With this relation, Eq.~\ref{current_locking5} is transformed to
\begin{equation}
\begin{aligned}
\frac{d\phi}{dt}&=-(\omega_0+Np)-\gamma\sigma^{\prime} J\epsilon \frac{1}{4\sqrt{p}}\sin(\psi)\\
\frac{d\psi}{dt}&=-\omega+\omega_e-N\delta p_s-\gamma\sigma^{\prime} J\epsilon \frac{1}{4\sqrt{p_0}}\sin(\psi)\\
&=\delta\omega_e-\nu\frac{\gamma\sigma^{\prime} J}{4\sqrt{p_0}}\epsilon\cos(\psi)-\gamma\sigma^{\prime} J\epsilon \frac{1}{4\sqrt{p_0}}\sin(\psi)\\
\end{aligned}
\label{current_locking7}
\end{equation}
where $\omega$ is the free-running auto-oscillation, and $\omega_e$ is the oscillation frequency of the external RF current. $\delta \omega_e = \omega_e-\omega$ is the detuning which measures the frequency difference between the free-running spin torque oscillator frequency $\omega$ and the external RF current frequency $\omega_e$. Eq.~\ref{current_locking7} is rearranged as
\begin{equation}
\frac{d\psi}{dt}=\delta\omega_e+\Delta\Omega\sin(\psi+\psi_0)
\end{equation}
where
\begin{equation}
\Delta\Omega = \frac{\gamma\sigma^{\prime} J\epsilon}{4\sqrt{p_0}}\sqrt{1+\nu^2}
\label{current_locking8}
\end{equation}

For stable phase locking, $d\psi/dt$ should be zero. Thus $\Delta\Omega$ is the phase locking range and $\psi_0$ is the phase difference between forced spin torque oscillator and injected RF current
\begin{equation}
\psi_0 = -\text{arctan}(\nu)
\end{equation}

If the STT term is expanded to the order of $p^2$, the locking range $\Delta\Omega$ is
\begin{equation}
\Delta\Omega = \frac{\gamma\sigma^{\prime} J\epsilon}{8\sqrt{p_0}}\sqrt{(2-3p_0)^2+[\nu(\frac{1}{2}+\lambda^2\sqrt{p_0(1-p_0)})(2-3p_0)]^2}
\label{current_locking9}
\end{equation}
and the phase difference is
\begin{equation}
\psi_0 = -\text{arctan}(\nu(\frac{1}{2}+\lambda^2\sqrt{p_0(1-p_0)})(2-3p_0))
\end{equation}

\begin{figure}
\includegraphics[width=0.45\textwidth]{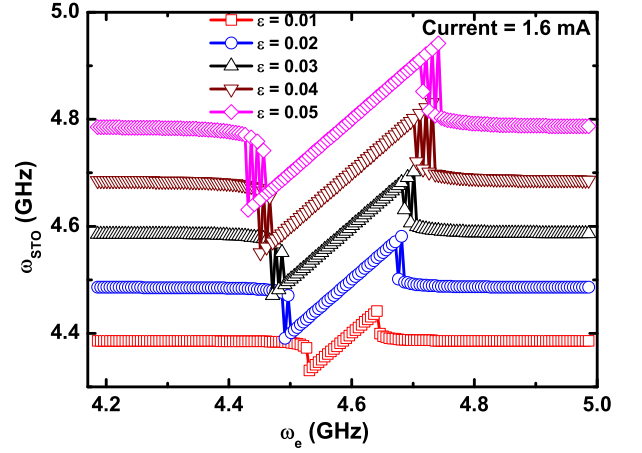}
\caption{\label{CLocking} Oscillation frequency in the phase locking state for injected DC current is 1.6 mA as a function of the external RF current frequency $\omega_e$. The macro simulation results are shown for RF/DC current ratio from 0.01 to 0.05. Plots are vertically offset for better readability.}
\end{figure}

\begin{figure}
\includegraphics[width=0.45\textwidth]{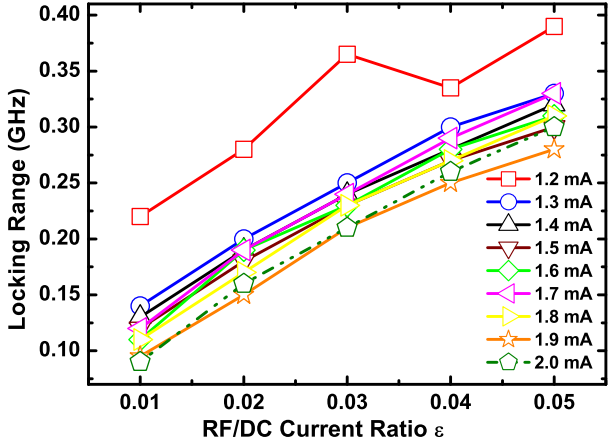}
\caption{\label{CLocking2} The locking range extracted from macro simulation results for varied RF/DC current ratio when injected DC current changes from 1.2 mA to 2.0 mA with step of 0.1 mA, respectively.}
\end{figure}

The phase locking phenomenon for spin torque oscillator with external injected RF current is simulated with macro-magnetic model, and the results are shown in Fig.~\ref{CLocking}. The DC current is 1.6 mA and the current amplitude ratio $\epsilon$ between RF current and DC current varies from 0.01 to 0.05. It is observed that the larger $\epsilon$ is, the larger the locking range is. The locking range for different injected DC current is extracted and plotted in Fig.~\ref{CLocking2}.

\begin{figure}
\includegraphics[width=0.45\textwidth]{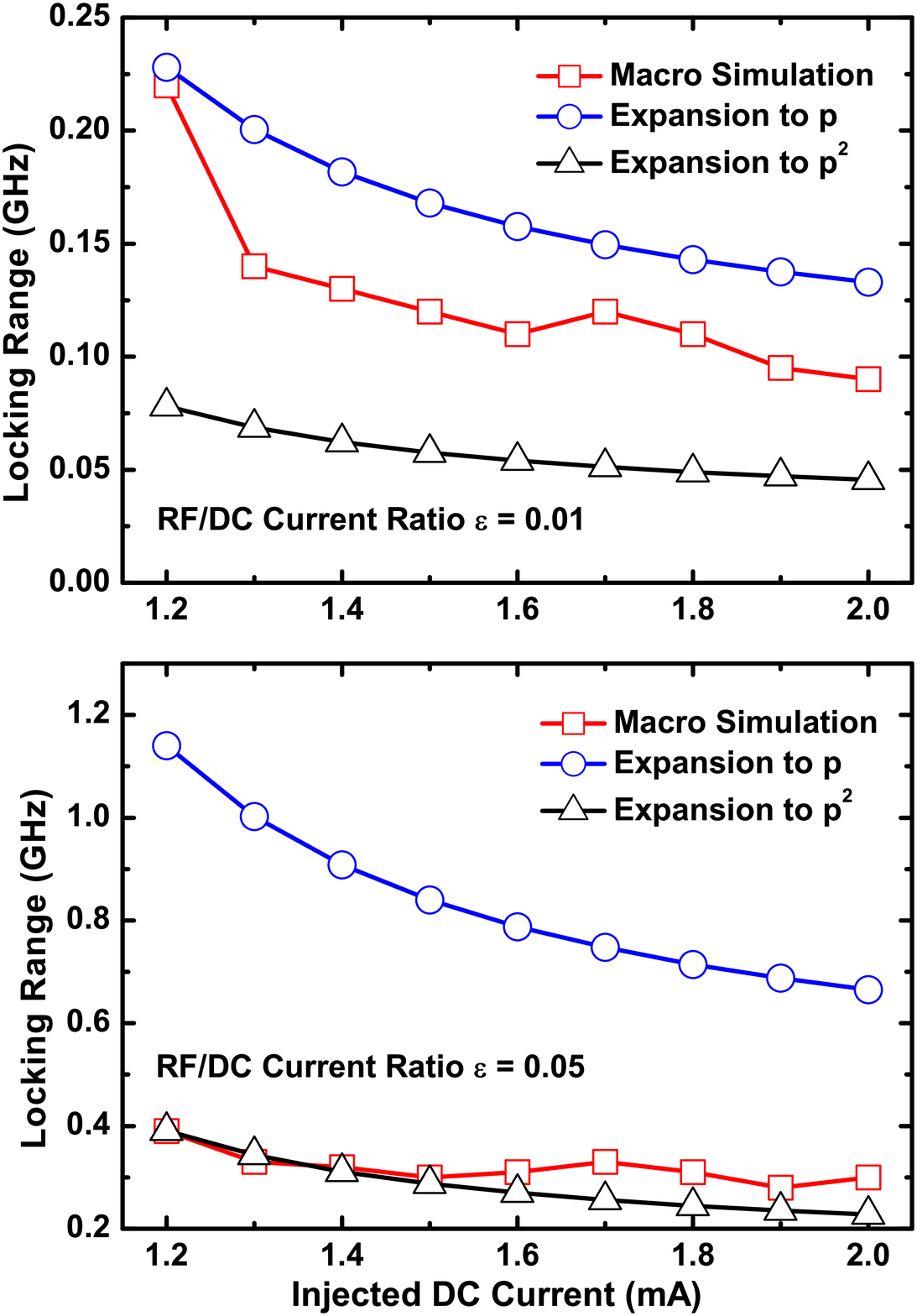}
\caption{\label{Fig7} Locking range when RF/DC current ratio $\epsilon = 0.01$ and $0.05$ for results extracted from macro simulation and that calculated by derived equation which corresponds to expansion to $p$ and $p^2$, respectively.}
\end{figure}

For comparing the macro simulation results and the analytical results, we plotted locking range extracted from macro simulation and computed with Eq.~\ref{current_locking8} and \ref{current_locking9} for varied DC current of 1.2 to 2.0 mA and current amplitude ratio $\epsilon$ of 0.01 to 0.05 in Fig.~\ref{Fig7}. It is can be seen that Eq.~\ref{current_locking8} over estimates the locking range while Eq.~\ref{current_locking9} under estimates the locking range. We can get similar observation from Fig.~\ref{fig1} where expansion to the order of $p$ under estimates the oscillation frequency and expansion to the order of $p^2$ over estimates the oscillation frequency, and expansion to the order of $p^3$ can give accurate estimations. Also we can observe that when $\epsilon = 0.01$, expansion to the order of $p$ can give a closer estimation for the locking range comparing with expansion to the order of $p^2$ while when $\epsilon = 0.05$, expansion to the order of $p^2$ can give very accurate results. For $\epsilon = 0.05$, the estimated locking range by Eq.~\ref{current_locking9} deviates more significantly when injected DC current is larger. This can be explained that the oscillation power $p$ increases with large injected DC current. So higher order expansion is needed for larger DC current.

\subsection{Theretical Analysis on Voltage Locking of VCSO}
\begin{figure}
\includegraphics[width=0.45\textwidth]{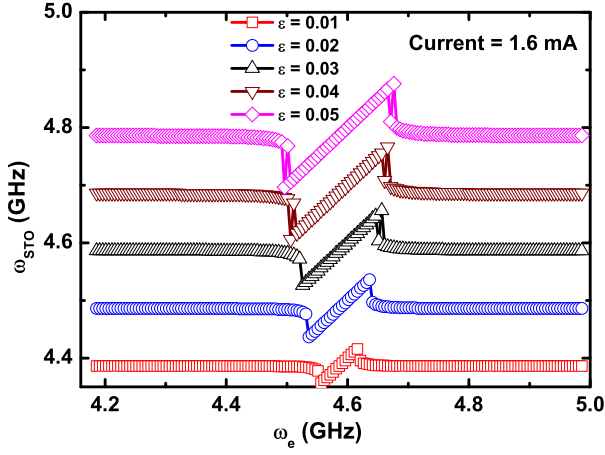}
\caption{\label{VLocking} Oscillation frequency in the phase locking state for injected DC current is 1.6 mA as a function of the external RF voltage frequency $\omega_e$. The macro simulation results are shown for RF/Reference VCMA voltage ratio from 0.01 to 0.05. Plots are vertically offset for better readability.}
\end{figure}

\begin{figure}
\includegraphics[width=0.45\textwidth]{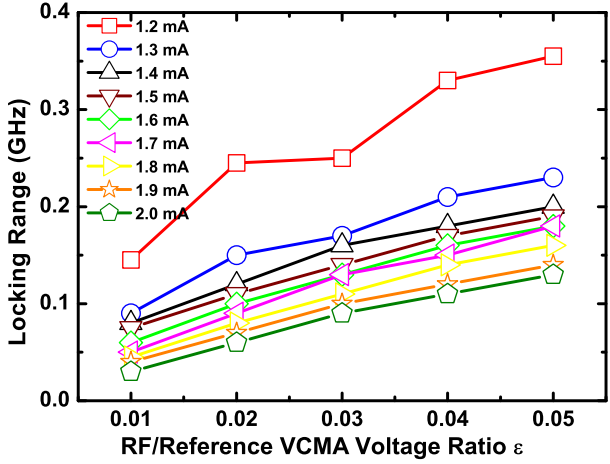}
\caption{\label{VLocking2} The locking range extracted from macro simulation results for varied RF/Reference VCMA voltage ratio when injected DC current changes from 1.2 mA to 2.0 mA with step of 0.1 mA, respectively.}
\end{figure}

The phase locking phenomenon of the spin torque oscillator is not novel. There are plenty reports in the literature about injection locking. However, since the applied VCMA voltage can tune the oscillation frequency of the VCSOs, a question will naturally be raised: will a RF VCMA voltage lead to phase locking? However, we followed the steps shown in Sec.~\ref{current_section} and surprisingly we find there is no coupling factors for resonant with external RF VCMA voltage. It implies RF VCMA voltage will not lock the phase of VCSO. But as plotted in Fig.~\ref{VLocking}, macro simulation results show that for injected DC current of 1.6 mA and no DC VCMA voltage, there does exist locking ranges when the frequency of the external RF voltages sweeps. Also it is seen that the locking range increases when the voltage amplitude ratio $\epsilon$ between RF VCMA voltage and reference VCMA voltage $V_\text{ref}$ increases from 0.01 to 0.05. The locking range with varied injected DC currents for voltage amplitude $\epsilon$ from 0.01 to 0.05 are extracted and plotted in Fig.~\ref{VLocking2}. It is worth noting that there are similar experimental results in the literature which reports a comparable voltage induced torque with spin transfer torque~\cite{VLocking_PRL}.

\begin{figure}
\includegraphics[width=0.5\textwidth]{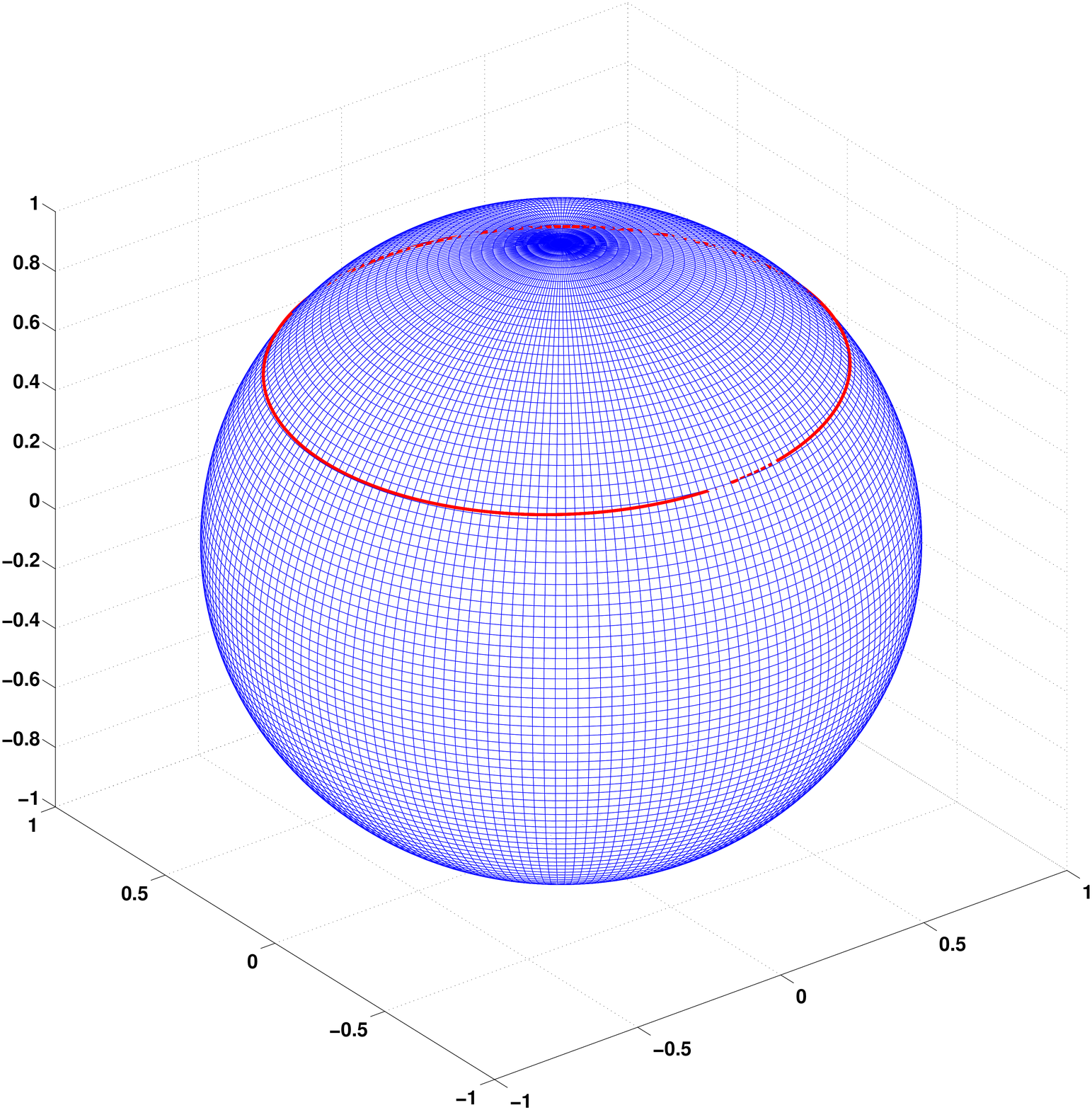}
\caption{\label{3D_rotate} 3D plot of the oscillating magnetization direction in the free layer. It can be observed that the trajactory of the oscillation is a little tilted away from the equal energy line, thus the oscillation axis is a little tilted away from z-axis.}
\end{figure}

The 3D plot for the trajectory of the oscillating magnetic moment of VCSO is shown in Fig.~\ref{3D_rotate}. Carefully observed, it can be seen that the oscillation orbit is slightly tilted away from the blue equal energy line. This means that the oscillation axis is not fully along the z axis, but is a little tilted away from the z axis. And this cant oscillation axis is the origin for voltage locking in VCSO.

Let us assume that the oscillation axis is tilted away from the z axis with angle $\theta$, and the magnetization direction in the energy minimum can be written as
\begin{equation}
m_{\theta} = m_x\sin(\theta)+m_z\cos(\theta)
\end{equation}
Thus the effective field $H_\text{eff}$ is got by
\begin{equation}
H_{\text{eff}} =  \left(\begin{array}{c}
H_{\text{appl}}\sin(\theta)+(H_k-4\pi M_s)\sin(\theta)m_{\theta}\\
0\\
H_{\text{appl}}\cos(\theta)+(H_k-4\pi M_s)\cos(\theta)m_{\theta}\\
\end{array}
\right)
\end{equation}
The precessional term $-\gamma(m\times H_\text{eff})$ is derived as
\begin{widetext} 
\begin{eqnarray} 
\begin{aligned}
\left.\frac{\partial a}{\partial t}\right|_\text{Precession} = & -i\gamma \left[\begin{array}{c}
H_{\text{appl}}+(H_k-4\pi M_s)a\cos(\theta)\\
-2(H_k-4\pi M_s)pa\cos(\theta)\\
\end{array}
\right]\cos(\theta)\\
&-i\gamma\frac{a^2+3p-2}{4\sqrt{1-p}}H_{\text{appl}}\sin(\theta)\\
&-i\gamma\frac{a^2+3p-2}{4}(a+a^*)(H_k-4\pi M_s)\sin(\theta)^2\\
&-i\gamma\frac{a(5a+11a^*-6pa)}{4\sqrt{1-p}}(H_k-4\pi M_s)\sin(\theta)\cos(\theta)\\
&+i\gamma\frac{(1+5p^2)}{2\sqrt{1-p}}(H_k-4\pi M_s)\sin(\theta)\cos(\theta)\\
\end{aligned}
\end{eqnarray} 
\label{voltage_locking1}
\end{widetext}
We considering this is no DC VCMA voltage and only RF VCMA voltage, the PMA field $H_\text{appl}$ reads
\begin{equation}
H_\text{appl}=(H_k-4\pi M_s)(1+\epsilon\cos(\phi_e))
\end{equation}
where $\epsilon$ is the voltage amplitude ratio of RF VCMA voltage and reference VCMA voltage $V_\text{ref}$, and $\phi_e$ is the phase of external RF VCMA voltage. Substituting $H_\text{appl}$ back into Eq.~\ref{voltage_locking1}, considering $\theta\ll 1$ and doing rearrangement we can get
\begin{equation}
\begin{aligned}
\left.\frac{\partial a}{\partial t}\right|_\text{Precession} = & -i\gamma \left[\begin{array}{c}
(H_{\text{appl}}+H_k-4\pi M_s)a\\
-2(H_k-4\pi M_s)pa\\
\end{array}
\right]\\
&-i\gamma\frac{5a^2+11p-6pa^2-10p^2-2}{4}*\\
&(H_k-4\pi M_s)\epsilon\cos(\phi_e)\sin(\theta)\\
\end{aligned}
\label{voltage_locking2}
\end{equation}
Keep in mind that in Eq.~\ref{voltage_locking2} only conservative term and resonating term with RF VCMA voltage are written.

Multiple $a^*$ in the left of Eq.~\ref{voltage_locking2}
\begin{widetext} 
\begin{eqnarray} 
\begin{aligned}
a^*\left(\frac{da}{dt}\right)_\text{Precession} = & -i\omega(p)p\\
&-i\gamma\frac{5pa+11pa^*-6p^2a-10p^2a^*-2a^*}{4}*\\
&(H_k-4\pi M_s)\epsilon\cos(\phi_e)\sin(\theta)\\
= & -i\omega(p)p\\
&-i\gamma\left(\frac{(\cos(\phi)+i\sin(\phi))(5p-6p^2)}{4}\right.\\
&\left.+\frac{(\cos(\phi)-i\sin(\phi))(11p-10p^2-2)}{4}\right)*\\
&(H_k-4\pi M_s)\epsilon\cos(\phi_e)\sin(\theta)\sqrt{p}\\
\end{aligned} 
\end{eqnarray} 
\end{widetext}
Further transform into oscillation power $p$ and oscillation phase $\phi$ if expanded to the order of $p^2$
\begin{equation}
\begin{aligned}
\frac{dp}{dt}=&-2(\Gamma_+(p)-\Gamma_-(p))p+(H_k-4\pi M_s)\epsilon\gamma*\\
&\frac{-6p+4p^2+2}{4}\sin(\psi)\sqrt{p}\sin(\theta)\\
\frac{d\phi}{dt}=&-(\omega_0+Np)-(H_k-4\pi M_s)\epsilon\gamma*\\
&\frac{16p-16p^2-2}{8\sqrt{p}}\cos(\psi)\sin(\theta)\\
\end{aligned}
\end{equation}
and expanded to the order of $p$
\begin{equation}
\begin{aligned}
\frac{dp}{dt}=&-2(\Gamma_+(p)-\Gamma_-(p))p+\\
&(H_k-4\pi M_s)\epsilon\gamma\frac{\sqrt{p}}{2}\sin(\psi)\sin(\theta)\\
\frac{d\phi}{dt}=&-(\omega_0+Np)+\\
&(H_k-4\pi M_s)\epsilon\gamma\frac{1}{4\sqrt{p}}\cos(\psi)\sin(\theta)\\
\end{aligned}
\label{voltage_locking3}
\end{equation}

The equation for the small power deviation $\delta p_s$ caused by forced external RF VCMA voltage is
\begin{equation}
0=-2\Gamma_p(p_0)\delta p_s+(H_k-4\pi M_s)\epsilon\gamma\frac{\sqrt{p_0}}{2}\sin(\psi)\sin(\theta)\\
\end{equation}
and $\delta p_s$ can be calculated as
\begin{equation}
\frac{\delta p_s}{p_0} = \frac{(H_k-4\pi M_s)\epsilon\gamma\sin(\psi)\sin(\theta)}{4\Gamma_p(p_0)\sqrt{p_0}}
\end{equation}

Using the same technology as the locking range derivation for injected RF current, we can get
\begin{equation}
\begin{aligned}
\frac{d\phi}{dt}=&-(\omega_0+Np)+\\
&(H_k-4\pi M_s)\epsilon\gamma\frac{1}{4\sqrt{p_0}}\cos(\psi)\sin(\theta)\\
\frac{d\psi}{dt}&=-\omega+\omega_e-N\delta p_s+\\
&(H_k-4\pi M_s)\epsilon\gamma\frac{1}{4\sqrt{p_0}}\cos(\psi)\sin(\theta)\\
&=\delta\omega_e-\nu\frac{(H_k-4\pi M_s)\epsilon\gamma\sin(\psi)\sin(\theta)}{4\sqrt{p_0}}\\
&+(H_k-4\pi M_s)\epsilon\gamma\frac{1}{4\sqrt{p_0}}\cos(\psi)\sin(\theta)\\
\end{aligned}
\end{equation}
Finally we can wirte
\begin{equation}
\frac{d\psi}{dt}=\delta\omega_e+\Delta\Omega\sin(\psi+\psi_0)
\end{equation}
where the locking range for voltage locking is written for expansion to the order of $p$
\begin{equation}
\Delta\Omega = (H_k-4\pi M_s)\epsilon\gamma\sin(\theta)\frac{1}{4\sqrt{p_0}}\sqrt{1+\nu^2}\\
\label{voltage_locking4}
\end{equation}
and the phase difference is written
\begin{equation}
\psi_0 = \pi-\text{arctan}(\nu)
\end{equation}
For the expansion to the order of $p^2$, we got
\begin{equation}
\begin{aligned}
\Delta\Omega =& (H_k-4\pi M_s)\epsilon\gamma\sin(\theta)\frac{1}{8\sqrt{p_0}}*\\
&\sqrt{(16p-16p^2-2)^2+\nu^2(-6p+4p^2+2)^2}\\
\end{aligned}
\label{voltage_locking5}
\end{equation}

\begin{figure}
\includegraphics[width=0.45\textwidth]{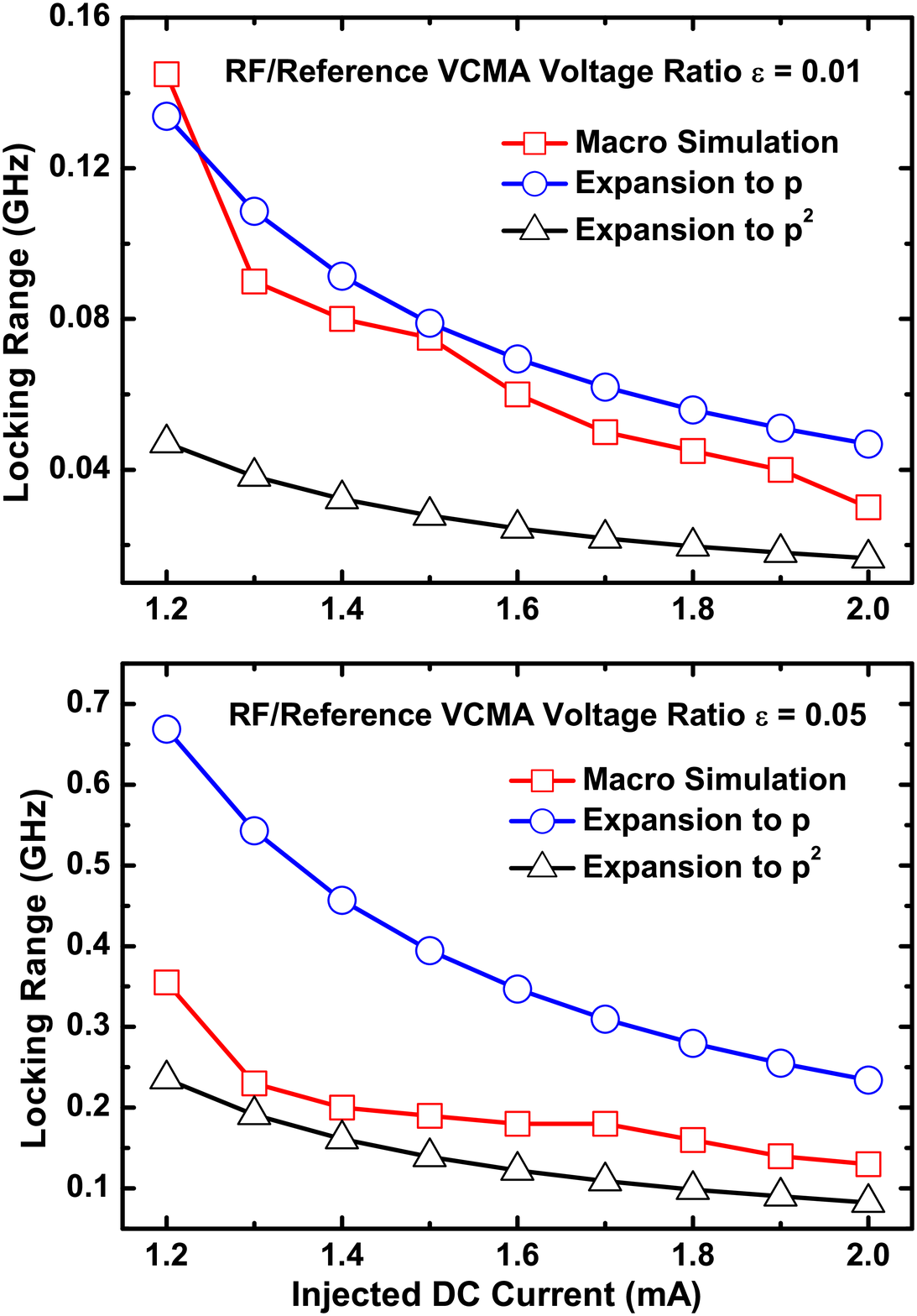}
\caption{\label{Fig10} Locking range when RF/Reference VCMA voltage ratio $\epsilon = 0.01$ and $0.05$ for results extracted from macro simulation and that calculated by derived equation which corresponds to expansion to $p$ and $p^2$, respectively.}
\end{figure}

The voltage locking range for varied injected DC current and no DC VCMA voltage are shown in Fig.~\ref{Fig10} for voltage amplitude ratio $\epsilon$ of 0.01 and 0.05 respectively. In Fig.~\ref{Fig10}, the macro-magnetic simulation results are compared with Eq.~\ref{voltage_locking4} for expansion to the order of $p$ and Eq.~\ref{voltage_locking5} for expansion to the order of $p^2$. Similar to the observation for current locking in Fig.~\ref{Fig7}, Eq.~\ref{voltage_locking4} can give a good estimation for $\epsilon = 0.01$ and Eq.~\ref{voltage_locking5} can give a good estimation for $\epsilon = 0.05$. This results are solid proof for the correctness of our theoretical derivation.

\begin{figure}
\includegraphics[width=0.45\textwidth]{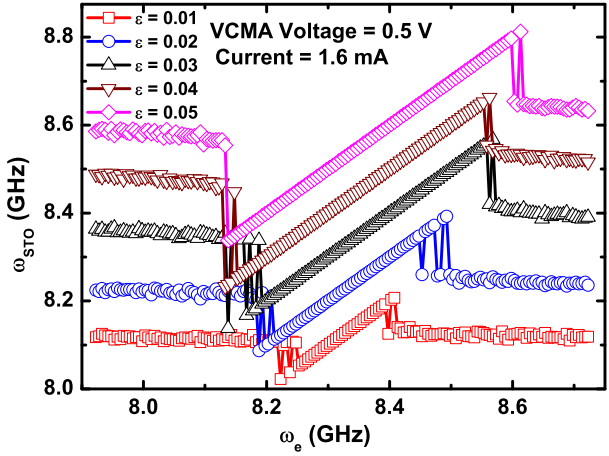}
\caption{\label{VoltageLocking} Oscillation frequency in the phase locking state for injected DC current is 1.6 mA and VCMA Voltage of 0.5 V as a function of the external RF voltage frequency $\omega_e$. The macro simulation results are shown for RF/Reference VCMA voltage ratio from 0.01 to 0.05. Plots are vertically offset for better readability.}
\end{figure}

\begin{figure}
\includegraphics[width=0.45\textwidth]{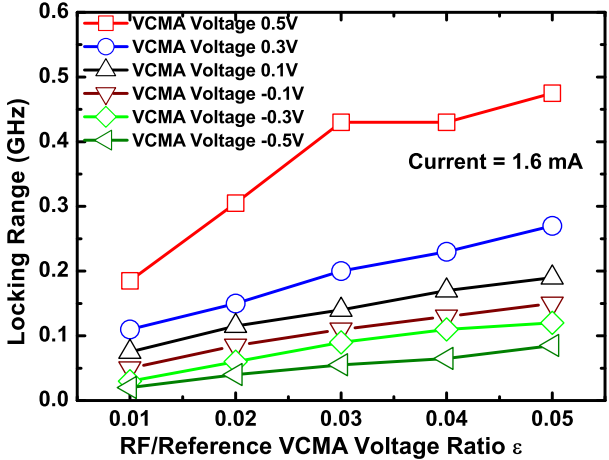}
\caption{\label{VoltageLocking2} The locking range extracted from macro simulation results for varied RF/Reference VCMA voltage ratio when injected DC current is 1.6 mA  and VCMA voltage varis from -0.5 V to 0.5 V with step of 0.2 V, respectively.}
\end{figure}

\begin{figure}
\includegraphics[width=0.45\textwidth]{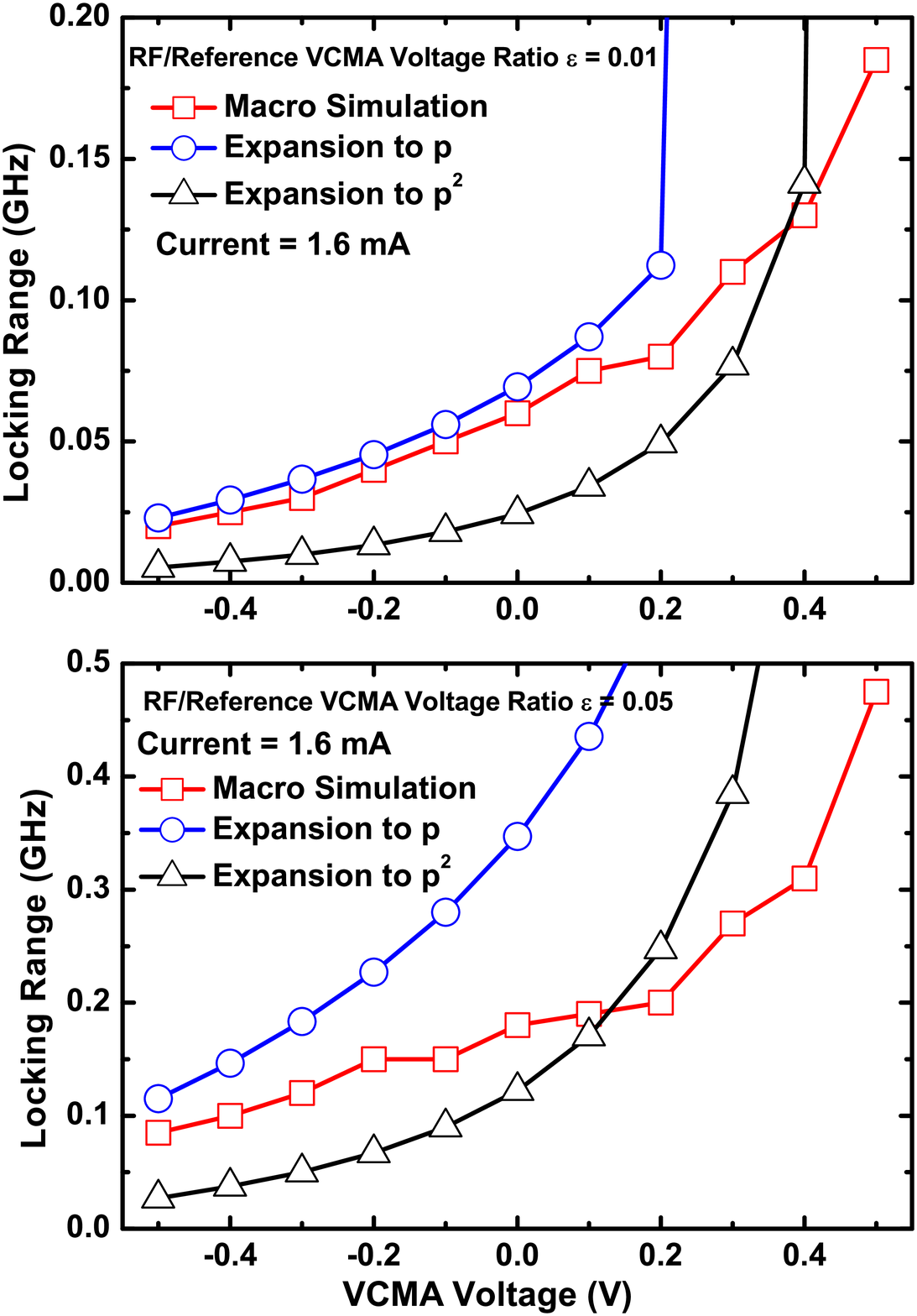}
\caption{\label{Fig15} Locking range dependence on VCMA voltage when RF/Reference VCMA voltage ratio $\epsilon = 0.01$ and $0.05$ for results extracted from macro simulation and that calculated by derived equation which corresponds to expansion to $p$ and $p^2$, respectively.}
\end{figure}

The case for non-zero DC VCMA voltage is also explored. Fig.~\ref{VoltageLocking} shows macro-magnetic simulations for DC current of 1.6 mA and DC VCMA voltage of 0.5 V when varied voltage amplitude $\epsilon$ when the frequency of the external RF VCMA voltage sweeps. And the voltage locking range for varied DC VCMA voltage extracted from macro-magnetic simulations are shown in Fig.~\ref{VoltageLocking2}. The comparison between macro-magnetic simulation results and analytical results is plotted in Fig.~\ref{Fig15}. When $\epsilon = 0.01$ and DC VCMA voltage is smaller than 0.4 V, Eq.~\ref{voltage_locking4} for expansion to the order of $p$ can give very good agreement. When $\epsilon = 0.05$ and DC VCMA voltage is smaller than 0.2 V, Eq.~\ref{voltage_locking5} for expansion to the order of $p^2$ can give very good agreement. However, when DC VCMA voltage is larger than 0.2 V, both of Eq.~\ref{voltage_locking4} and \ref{voltage_locking5} over estimate the voltage locking range. When DC VCMA voltage is larger than 0.2 V, the oscillation power $p$ will be small from Fig.~\ref{FvsV} and \ref{fig4} and $\nu$ drops dramatically. Thus maybe expansion to higher order of $p$ is required when DC VCMA voltage is large than 0.2 V.

\subsection{Expanding Locking Range by Enhanced VCMA effect}

Although STNO is very promising device to replace VCO based on semiconductor transistors, low emitted power and high phase noise hinders its practical use in modern communication system. However, both of the figure of merit for emitted power and phase noise can be dramatically improved by mutual synchronization of multiple STNOs. There are several synchronization mechanisms proposed in the literature: spin wave, magnetic field, dipolar effect as well as electrical connection. Among all of these means, the synchronization through electrical connection is most favorable for implemented on chip. As shown in Ref.~17 only in vortex STNO the mutual synchronization through electrical connection is experimentally demonstrated. This is because of high emitted power of vortex STNO can overcome the random walking phase error induced by thermal fluctuation.

The phase locking phenomenon caused by injected RF current and applied RF voltage is closely related to the mutual synchronization through electrical connection. The wider the locking range is, the easier the mutual synchronization is. So it will have significant impact on mutual synchronization by electrical connection if the locking range can be expanded. From the previous sections about the theoretical and numerical results of locking range for current and voltage locking, we can see that the locking range for current and voltage locking are comparable. However, there are two methods which can significantly expand the locking range of voltage locking.

1) From Eq.~\ref{voltage_locking4} and \ref{voltage_locking5}, the locking range for current locking $\Delta\Omega$ is directly proportional to $\sin(\theta)$. From Fig.~\ref{3D_rotate}, we can see that $\theta$ is very small and the oscillation axis is only a little bit tilted away from z axis. As pointed out in Ref.~35 and 37, if the field like torque of the spin polarized current taken into account, the oscillation axis will be tilted away in a large angle from the z axis. This will lead to a tremendously expanded locking range to facilitate easy mutual synchronization by electrical connection. However, the details analysis for mutual synchronization by voltage coupling should be carefully taken and is beyond the scope of this work. It will be our future work.

\begin{figure}
\includegraphics[width=0.5\textwidth]{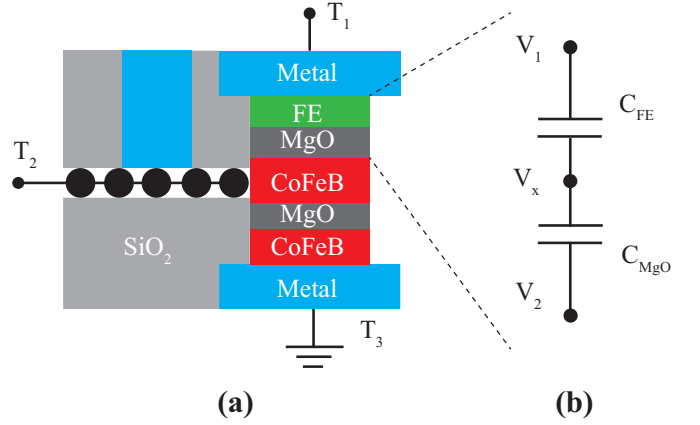}
\caption{\label{VCSO2} (a) The structure of the proposed Voltage Controlled Spin Oscillator (VCSO) with negative capacitance ferroelectric material to ampilify VCMA effect. (b) The simplified capacitance model for the calculatio of the VCMA amplification ratio.}
\end{figure}

2) In our previous study, we proposed that negative capacitance ferroelectric material can be used to amplify the VCMA effect for the first time~\cite{zeng1,zeng2}. As shown in Fig.\ref{VCSO2} (a), the structure we proposed to amplify VCMA effect just requires an additional ferroelectric layer. The principle for why the capacitance of the ferroelectric layer can be negative is proposed in Ref.~47, and it is very hot topic in the semiconductor device community to achieve steep slope transistors~\cite{NC1,NC2}. Recently there are two back to back papers published in Nature to report direct observation of negative capacitance phenomenon~\cite{NC3,NC4}. The simplified capacitance model for the proposed structure is shown in Fig.~\ref{VCSO2} (b). As shown, the internal voltage $V_x$ which is the effective voltage applied on MgO insulating layer to induce VCMA effect can be calculated as when $V_1 = 0$
\begin{equation}
V_x = \frac{V_2 C_{FE}}{C_{MgO}+C_{FE}}
\end{equation}
$V_x$ will be much larger than the external applied voltage $V_2$ if $C_{FE}+C_{MgO}\approx 0$. The amplification ratio of the VCMA effect is defined as
\begin{equation}
\frac{V_x}{V_2}=\frac{C_{tot}}{C_{MgO}} = \frac{C_{FE}}{C_{MgO}+C_{FE}}
\label{Ctot}
\end{equation}
In our previous study, we have proved that by this novel method, the VCMA effect can be amplified to tens or even hundreds of times. And based on the amplification of the VCMA effect, we have proposed memory devices as well as all spin logic device with features of ultra fast, ultra low power consumption and ultra low working voltage~\cite{zeng1,zeng2}. According to Eq.~\ref{voltage_locking4} and \ref{voltage_locking5}, the locking range for current locking $\Delta\Omega$ is directly proportional to $\epsilon$. $\epsilon$ is proportional to VCMA coefficient thus giant VCMA effect will lead to very large locking range. More details analysis on this kind of method is beyond the scope of this work and will be our next work.

\section{Conclusion}
In this work, we proposed a Voltage Controlled Spin Oscillator (VCSO). The VCMA effect is employed to play as an additional degree of freedom to tune oscillation frequency of the VCSO. We also proposed that applying a RF voltage can lead to novel phase locking phenomenon as well as current locking in our VCSO. Both of the frequency modulation and current and voltage locking are theoretically investigated in the framework of Nonlinear Auto-oscillation theory. The analytical results agree very well with the numerical simulation results. In the end, two kinds of method which can expand the locking range for voltage locking are proposed. Especially we proposed that by integrating an additional ferroelectric layer, the VCMA effect can be amplified thus lead to largely expanded locking range as well as possible easy mutual synchronization in uniform precession STNOs by voltage coupling through electrical connection.

\begin{acknowledgments}
The authors wish to acknowledge the support from the National Natural Science Foundation of China under Grants 61471015, 61504006 and 51602013, International mobility project under Grant B16001, National Key Technology Program of China under Grant 2017ZX01032101, National Postdoctoral Program for Innovation Talents under Grant BX20180028 and China Postdoctoral Science Foundation funded project under Grant 2018M641153.
\end{acknowledgments}

\appendix
\section{Virtual Perpendicular System for Fast Theoretical Derivation\label{VPS}}

As shown by Eq.~\ref{p2} and \ref{p3}, the calculation of the STT term expressed with $a$ involved Taylor expansion and merging of  items with the same order. This process is tedious, cumbersome and easy to make mistakes. Here we provide a novel method which can get the conservative terms of the STT term in a very convenient way.

Let us start from a STNO which has both of the free layer and reference layer in the perpendicular direction. The STT terms of such configuration can be calculated as
\begin{equation}
\left.\frac{\partial a}{\partial t}\right|_\text{STT} = \sigma J(1-p)a
\label{PS}
\end{equation}
where $\sigma$ is the spin torque strength in this STNO
\begin{equation}
\begin{aligned}
\sigma& = \frac{\gamma\hbar\eta}{2eM_s Sd}\\
\end{aligned}
\end{equation}
From Eq.~\ref{PS}, we can see that in such a STNO there is only conservative term but no resonant term. It is can be concluded that injection locking by RF current is absent in such system.

By using the LLG equation, the work done by the spin torque $W$ is given by
\begin{equation}
\begin{aligned}
W = &\gamma M_s \sigma[e_p\cdot H_\text{eff}-(m\cdot e_p)(m\cdot H_\text{eff})\\
&-\alpha e_p\cdot(m\times H_\text{eff})]\\
\end{aligned}
\label{W}
\end{equation}

$W$ for the STNO with both of the free layer and reference layer in the perpendicular direction (we call it perpendicular system) is very easy to calculate
\begin{equation}
W_\text{PS} = \frac{\gamma M_s \hbar \eta J}{2eM_sSd}(H_\text{appl}+(H_k-4\pi M_s)\cos(\theta))\sin(\theta)^2
\label{WPS}
\end{equation}
where a constant tilted angle $\theta$ for a steady precession around the z-axis is assuming.

While in Ref.~35, $W$ for the STNO with the same configuration as our proposed VCSO (we can call it Virtual Perpendicular System, and the reason will be explained in the following) is given by~\cite{Kubota4}
\begin{equation}
\begin{aligned}
W_\text{VPS} = &\frac{\gamma M_s \hbar \eta J}{2e\lambda M_sSd}\left(\frac{1}{\sqrt{1-\lambda^2\sin(\theta)^2}}-1\right)\\
&[H_\text{appl}+(H_k-4\pi M_s)\cos(\theta)]\cos(\theta)\\
\end{aligned}
\label{WVPS}
\end{equation}
where a constant tilted angle $\theta$ for a steady precession around the z-axis is assuming.

Actually, calculation of the work done by spin torque or damping term is the typical operation in energy balance theory. Since $W_\text{PS}$ or $W_\text{VPS}$ is integration over one period of the oscillation orbit, it should be the same as calculation of the conservative terms like Eq.~\ref{p2}, \ref{p3} or Eq.~\ref{PS}. Thus we postulate the following equation should be established
\begin{equation}
\left.\frac{\partial a}{\partial t}\right|_\text{VPS} = \frac{W_\text{VPS}}{W_\text{PS}}\cdot\left.\frac{\partial a}{\partial t}\right|_\text{PS}
\end{equation}
Then we can get
\begin{equation}
\left.\frac{\partial a}{\partial t}\right|_\text{VPS} = (1-p)\left(\frac{1}{\sqrt{1-\lambda^2\sin(\theta)^2}}-1\right)\frac{\cos(\theta)}{\lambda\sin(\theta)^2}
\end{equation}
Substituting $p=(1+\cos(\theta))/2$
\begin{equation}
\begin{aligned}
\left.\frac{\partial a}{\partial t}\right|_\text{VPS} = &(1-p)\left(\frac{1}{\sqrt{1-\lambda^2(1-(1-2p)(1-2p))}}-1\right)*\\
&\frac{1-2p}{\lambda(1-(1-2p)(1-2p))}\\
\end{aligned}
\label{VPS2}
\end{equation}
Taylor expansion of Eq.~\ref{VPS2} to the order of $p^2$, Eq.~\ref{p2} can be easily got while Taylor expansion of Eq.~\ref{VPS2} to the order of $p^3$, Eq.~\ref{p3} can be easily got. Any we can calculate STT term to any order of $p$ as we want.

Although VCSO is not perpendicular system which should have perpendicular free layer and reference layer, we successfully make a connection between the work done by spin torque and the STT term. Thus we can calculate the STT term from that of the perpendicular system. In other words, the VCSO can be treated as a Virtual Perpendicular System. It is worth noting that such connection is general relation, thus our VPS method can be applied to other STNOs with different configurations. It is a general fast theoretical derivation method.

\bibliography{VCSO}

\end{document}